\documentclass[prd,aps,floats,floatfix,superscriptaddress,preprintnumbers,
showpacs,eqsecnum,nofootinbib,twocolumn]{revtex4}
\usepackage{latexsym,array,theorem,mathrsfs,bm,float}

\usepackage{psfrag}
\usepackage{amsfonts,amsmath,amssymb,latexsym,array,afterpage,theorem,mathrsfs,bm,float,epsfig,color,graphicx,tabularx,here,multirow}
\newcommand{\nn}{\nonumber \\}
\newcommand{\bea}{\begin{eqnarray}}
\newcommand{\ena}{\end{eqnarray}}
\newcommand{\beann}{\begin{eqnarray*}}
\newcommand{\enann}{\end{eqnarray*}}

\newcommand{\ma}[1]{\mbox{$\mathcal{#1}$}}

\newcommand{\calhR}[1]{\raisebox{2ex}{\tiny ({\em h})}\hspace{-0.8em}{\ma R}}

\newcommand{\mpl}{M_{\mathrm{PL}}}

\begin{document}

\title{
Cosmological Dynamics of Cuscuta-Galileon Gravity
}


\author{Sirachak {\sc Panpanich}}
\email{sirachakp-at-aoni.waseda.jp}
\address{Department of Pure and Applied Physics, Graduate School of Advanced Science and Engineering, Waseda University, 
Okubo 3-4-1, Shinjuku, Tokyo 169-8555, Japan}
\author{Kei-ichi {\sc Maeda}}
\email{maeda-at-waseda.jp}
\address{Department of Pure and Applied Physics, Graduate School of Advanced Science and Engineering, Waseda University, 
Okubo 3-4-1, Shinjuku, Tokyo 169-8555, Japan}


\date{\today}

\begin{abstract}

We study cosmological dynamics of the cuscuta-galileon gravity with a potential term by using the dynamical system approach. This model is galileon generalization of the cuscuton gravity where we add a potential term to the theory in order to obtain the radiation and matter dominated eras. The exponential potential can provide the sequence of the thermal history of the Universe correctly, i.e. starting from radiation dominance, passing through matter dominant era, and then approaching de Sitter expansion stage. 
 This model has no ghosts and the Laplacian instability for both scalar and tensor perturbations. 
We also discuss the observational constraints on the  model parameters.
It turns out  that the model actually has three degrees of freedom unlike the original cuscuton theory. 

\end{abstract}


\maketitle

\section{Introduction}

Many modified gravity models require additional degrees of freedom (d.o.f.) besides two tensor gravitational degrees of freedom to explain an accelerated expansion of the Universe \cite{Riess:1998cb,Perlmutter:1998np}. For example, Horndeski theories \cite{Horndeski:1974wa,Deffayet:2011gz,Kobayashi:2011nu} using a scalar field have 
three d.o.f., generalized Proca theories \cite{Heisenberg:2014rta} using a vector field have five d.o.f., and massive gravity using a massive tensor field has five d.o.f in the case of de Rham-Gabadadze-Tolley massive gravity \cite{deRham:2010ik,deRham:2010kj}. However, until now, a fifth force or deviation from General Relativity in the solar system scale has not been detected \cite{Will:2014kxa}. Therefore, they require screening mechanisms to hide their additional degrees of freedom \cite{Khoury:2003aq,Khoury:2003rn,Hinterbichler:2010es,Brax:2010gi,Burrage:2014uwa,Panpanich:2019rij,Brax:2012jr,Babichev:2009ee,Babichev:2013usa,Vainshtein:1972sx,Nicolis:2008in}.

Recently there is development on modification of the gravitational theories which propagate only two gravitational degrees of freedom. 
The cuscuton gravity model was first proposed in  \cite{Afshordi:2006ad,Afshordi:2007yx,Afshordi:2009tt}, which can be regarded as the low-energy Horava-Lifshitz theory\cite{Afshordi:2009tt}.
Some extension was also found in
 minimally modified gravity (MMG) \cite{Lin:2017oow,Aoki:2018zcv,Aoki:2018brq,Mukohyama:2019unx,DeFelice:2020eju,Aoki:2020oqc} and extended cuscuton gravity \cite{Iyonaga:2018vnu,Iyonaga:2020bmm}. The minimally modified gravity is a construction of Hamiltonian of the gravitational theory which provides only two d.o.f., while the extended cuscuton is a generalization of an original cuscuton theory  in the context of the beyond Horndeski theories \cite{Gleyzes:2014dya}.
 In these models
  the scalar field turns out to be  nondynamical
  because  either the second-order time derivatives of a scalar field are absent in the equations of motion (MMG), or  we can eliminate them 
 after linear combination of the equations of motion and the Friedmann equation (extended cuscuton). 

Besides above two classes of theories, the cuscuta-galileon gravity  which is a simple galileon generalization of the original cuscuton gravity, was proposed \cite{deRham:2016ged}. 
 In the original cuscuton gravity, there exists  
a caustic singularity, which shows lacking predictability.
Hence adding a galileon-like kinetic term in the original cuscuton theory,
they discuss a simple extended model (the cuscuta-galileon theory), which
 can avoid the formation of  caustic singularities in flat space-time \cite{deRham:2016ged}.

In this work we investigate cosmological dynamics of the cuscuta-galileon gravity. 
We include a potential term because
without a potential term, such a model does not provide a viable cosmological model just as the same as the original cuscuton theory
 \cite{Afshordi:2006ad,Afshordi:2007yx}. 
In fact we  find that 
radiation dominant and matter dominant eras do not exist as we will show in Appendix \ref{cuscutawithoutpotential}. 
In the original cuscuton theory, adding a quadratic potential,
 we obtain the $\Lambda$CDM model.
Therefore, we imitate this idea by adding a potential term to the cuscuta-galileon action, and then investigate cosmological dynamics of the model with an appropriate potential term whether it provides 
a consistent cosmic evolution or not.

This paper is organized as follows. In section \ref{basiceq} we derive basic equations of the model. In section \ref{dynamical} we study the cosmological dynamics by using the dynamical system approach where we consider a scalar potential in two cases: an exponential potential and an inverse power-law potential. In section \ref{dof} we use the Hamiltonian formalism to investigate number degrees of freedom of the cuscuta-galileon model rigorously. 
It turns out that the cuscuta-galileon gravity in fact has three d.o.f. which leads to tendency that the model is not in a subclass of the extended cuscuton gravity, but rather in a subclass of the Horndeski theories. 
In section \ref{numer},
we solve autonomous equations of the model numerically and show evolution of density parameters and equation of state parameters. We check ghosts and Laplacian instabilities in section \ref{perturbations}. Lastly, section \ref{conclusions} is devoted to conclusions.


\section{Action and Basic equations}
\label{basiceq}

We start at an action of the cuscuta-galileon gravity as Ref. \cite{deRham:2016ged} in curved space-time with a potential term,
~~~~~~~~~~~~~~~~~~~~
\bea
S &=& \int d^4 x \sqrt{-g} \Big[ \frac{1}{2} \mpl^2 R + a_2 \sqrt{-X} + a_3 \ln \Big(-\frac{X}{\Lambda^4}\Big) \square \phi  \nn 
& & - V(\phi)\Big] + S_M (g_{\mu\nu}, \psi_M) \,, \label{action}
\ena
where $R$ is the Ricci scalar, $\mpl$ is the reduced Planck mass, $g$ is the determinant of the metric $g_{\mu\nu}$, and $\psi_M$ is a fermion field. $X$ is defined as $X \equiv g^{\mu\nu}\partial_{\mu} \phi \partial_{\mu} \phi$. $a_2$, $a_3$, and $\Lambda$ are constants with dimension of mass squared, mass, and mass, respectively. 
 We consider up to cubic order to satisfy the constraint from the gravitational waves observations, GW170817 \cite{Baker:2017hug,Creminelli:2017sry,Sakstein:2017xjx,Ezquiaga:2017ekz}, and  add the potential term in order to obtain radiation dominated and matter dominated eras according to the thermal history of the Universe. Without the potential term, the theory provides only the de Sitter expansion as shown in Appendix \ref{cuscutawithoutpotential}.

We consider the flat Friedmann-Lema\^{\i}tre-Robertson-Walker (FLRW) metric and 
 a homogeneous scalar field as
\bea
ds^2 = - N(t)^2 dt^2 + a(t)^2 \delta_{ij} dx^i dx^j \,, \quad  \phi = \phi(t) \,.
\ena
Substituting the metric into the above action, and then varying with respect to $N$, $a$, and $\phi$, after setting $N = 1$ we find
\bea
3\mpl^2 H^2 - \rho_m - \rho_r - V(\phi) + 6 a_3 H \dot \phi &=& 0 \,,\label{1stFriedmann} \\
3\mpl^2 H^2 + 2\mpl^2 \dot H + P_m + P_r - V(\phi) \nn + a_2 \lvert \dot \phi \rvert + 2a_3 \ddot \phi &=& 0 \,, \label{2ndFriedmann} \\
18 a_3 H^2 + 6 a_3 \dot H - V_{,\phi} - 3a_2 H {\rm sgn} (\dot\phi) &=& 0 \,, \label{EOM}
\ena
where $\rho_m$, $P_m$, $\rho_r$, and $P_r$ are densities and pressures of nonrelativistic matter (or matter for abbreviation) and radiation, respectively. $H$ is the Hubble parameter, an upper dot means the derivative with respect to time, $``~_{,\phi}"$ denotes the partial derivative with respect to $\phi$, and ${\rm sgn} (\dot\phi)$ is the sign of $\dot\phi$. Eqs. (\ref{1stFriedmann}) and  (\ref{2ndFriedmann}) are the Friedmann equations, and Eq. (\ref{EOM}) is the equation of motion of the scalar field. Although there is no second-order time derivatives of the scalar field in the equation of motion, there is ambiguity because these three basic equations are not independent as shown in Appendix \ref{independenteq} which allows us to write independent equations for $\ddot\phi$ and $H$. Therefore in order to know exact number degrees of freedom we have to perform the Hamiltonian analysis which will be given in section \ref{dof}.

From the Friedmann equations we can define density and pressure of the scalar field as 
\bea
\rho_{\phi} &=& V(\phi) - 6a_3 H \dot \phi \,, \\
P_{\phi} &=& a_2 \lvert \dot\phi \rvert + 2 a_3 \ddot\phi - V(\phi) \,.
\ena
Combination of the Friedmann equations and the equation of motion, we obtain the energy conservation equations: 
\bea
\dot\rho_m + 3H \rho_m &=& 0 \,, \\
\dot\rho_r + 4H \rho_r &=& 0 \,, \label{conservedradiation} \\
\dot\rho_{\phi} + 3H (\rho_{\phi} + P_{\phi}) &=& 0 \,,
\ena
where we assume that the nonrelativistic matter is pressureless, $P_m \approx 0$, while the pressure of radiation is $P_r = \rho_r / 3$. 

In the next section we will use dynamical system approach to study cosmological dynamics of the cuscuta-galileon gravity. 


\section{Dynamical System}
\label{dynamical}

\subsection{Autonomous Equations}

\allowdisplaybreaks

We introduce dimensionless variables as follows
\begin{gather}
x_1 \equiv \frac{V}{3\mpl^2 H^2} \,, \quad x_2 \equiv \frac{a_2 \lvert \dot\phi \rvert}{\mpl^2 H^2} \,, \quad x_3 \equiv \frac{2a_3 \dot\phi}{\mpl^2 H} \,, \nn
x_4 \equiv \frac{\rho_r}{3\mpl^2 H^2} \,, \quad \lambda \equiv \frac{\mpl^2 V_{,\phi}}{a_3 V} \,. \label{dimensionlessvariables}
\end{gather}
Thus the first Friedmann equation (\ref{1stFriedmann}) can be written as
\bea
\Omega_m = 1 - x_1 + x_3 - x_4 \,, \label{omegam}
\ena
where $\Omega_m \equiv \rho_m / 3\mpl^2 H^2$ is a density parameter of the nonrelativistic matter. This is a constraint equation where dynamics of the matter density parameter can be realized via variables $x_1$, $x_3$, and $x_4$. Also, density parameters of the radiation and the scalar field are
\bea
\Omega_r \equiv \frac{\rho_r}{3\mpl^2 H^2} = x_4 \,, \quad \Omega_{\phi} = x_1 - x_3 \,. \label{omegarandphi}
\ena
Taking derivative with respect to the e-foldings number, $N \equiv \ln a$, we find a set of autonomous equations:
\bea
\frac{dx_1}{dN} &=& \frac{1}{2} \lambda x_1 x_3 - 2x_1 \frac{\dot H}{H^2} \,, \label{auto1} \\
\frac{dx_2}{dN} &=& \frac{x_2}{x_3} \Big( 3 - x_4 + 3 x_1 - \lambda x_1 - x_2 - \frac{2x_2}{x_3} \Big) \nn & & - 2 x_2 \frac{\dot H}{H^2} \,, \label{auto2} \\
\frac{dx_3}{dN} &=& 3 - x_4 + 3 x_1 - \lambda x_1 - x_2 - \frac{2x_2}{x_3} - x_3 \frac{\dot H}{H^2} \,, \label{auto3} \quad \\
\frac{dx_4}{dN} &=& -4 x_4 -2 x_4 \frac{\dot H}{H^2} \,, \label{auto4} \\
\frac{d\lambda}{dN} &=& \frac{1}{2}\lambda^2 x_3 (\Gamma - 1) \,, \label{auto5}
\ena
and
\beann
\frac{\dot H}{H^2} = - 3 + \frac{1}{2} \lambda x_1 + \frac{x_2}{x_3} \,, \quad \Gamma \equiv \frac{V V_{,\phi\phi}}{V_{,\phi}^2} \,.
\enann
We have used the second Friedmann equation (\ref{2ndFriedmann}), the equation of motion (\ref{EOM}), and the continuity equation of radiation (\ref{conservedradiation}) to obtain these autonomous equations. 

Effective equation of state parameter and equation of state parameter of the scalar field are defined as
\bea
w_{\rm eff} &=& \frac{P_{\rm total}}{\rho_{\rm total}} = 1 - \frac{2x_2}{3x_3} - \frac{1}{3}\lambda x_1 \,, \label{weff} \\
w_{\phi} &=& \frac{P_{\phi}}{\rho_{\phi}} = \frac{\displaystyle{w_{\rm eff} - \frac{1}{3}x_4}}{x_1 - x_3}  \,. \label{wphi}
\ena
Next we will consider the cuscuta-galileon model in two cases: an exponential potential and an inverse power-law potential.

\subsection{Fixed Points}

\subsubsection{Exponential potential}
 
If the potential is an exponential form, the $\lambda$ is a constant. We then have only $4$ autonomous equations (\ref{auto1}) - (\ref{auto4}) with $4$ parameters. Integrating the definition of $\lambda$ in the Eq. (\ref{dimensionlessvariables}), we find
\bea
V(\phi) = V_0 e^{a_3 \lambda \phi / \mpl^2} \,,
\ena
where $V_0$ is a constant. Setting $dx_1/dN = dx_2/dN = dx_3/dN = dx_4/dN = 0$, we find five fixed points as Table \ref{fixedpoints}.

\begin{widetext}

\begin{table}[h]
\begin{center}
  \begin{tabular}{c c c c c c c c c c}
\hline 
\hline 
& & & & & & & & & \\[-.5em]
Fixed point& $x_1$ & $x_2$ & $x_3$ & $x_4$ & $\Omega_m$ & $\Omega_r$ & $\Omega_{\phi}$ & $w_{\phi}$ & $w_{\rm eff}$ 
\\[.5em]
\hline
& & & & & & & & & \\[-.5em]
(a) & $0$ & $-3$ & $-1$ & $0$ & $0$ & $0$ & $1$ & $-1$ & $-1$
\\[.5em]
& & & & & & & & & \\[-.5em]
(b) & $0$ & $0$ & $-1$ & $0$ & $0$ & $0$ & $1$ & $1$ & $1$
\\[.5em]
& & & & & & & & & \\[-.5em]
(c) & $\frac{2}{\lambda}$ & $0$ & $-\frac{8}{\lambda}$ & $1 - \frac{10}{\lambda}$ & $0$ & $1 - \frac{10}{\lambda}$ & $\frac{10}{\lambda}$ & $\frac{1}{3}$ & $\frac{1}{3}$
\\[.5em]
& & & & & & & & & \\[-.5em]
(d) & $\frac{3}{\lambda}$& $0$ & $-\frac{6}{\lambda}$ & $0$ & $1 - \frac{9}{\lambda}$ & $0$ & $\frac{9}{\lambda}$ & $0$ & $0$
\\[.5em]
& & & & & & & & & \\[-.5em]
(e) & $-1+\frac{12}{\lambda}$ & $0$ & $-2+\frac{12}{\lambda}$ & $0$ & $0$ & $0$ & $1$ & $-3 + \frac{\lambda}{3}$ & $-3 + \frac{\lambda}{3}$
\\[.5em]
\hline 
 \end{tabular}
    \caption{The fixed points, the density parameters, and the equation of state parameters of the cuscuta-galileon with the exponential potential.}
\label{fixedpoints}
\end{center}
\end{table}

\end{widetext}

In this work we are interested in the case $V(\phi) \geq 0$ (i.e., $V_0 > 0$) and the expanding universe, $H > 0$, thus the $x_1 \geq 0$. The $x_2$ and the $x_3$ can be positive or negative values depending on the signs of $a_2$, $a_3$, and $\dot\phi$. The $x_4 \geq 0$ because $\rho_r \geq 0$. These conditions lead to constraints on $\lambda$ of some fixed points. The fixed point (c) requires $\lambda \geq 10$ to satisfy $x_4 \geq 0$, whereas the fixed point (d) and (e) require $\lambda > 0$ and $\lambda \leq 12$ to satisfy $x_1 \geq 0$, respectively.

Considering the equation of state parameters and the density parameters, we find that only the fixed point (c) can be the radiation dominated epoch.
Rigorously, this fixed point is the $\phi$-radiation dominated epoch because 
 $\Omega_{\phi}$ does not vanish but keeps constant. 
The energy density of the scalar field decreases in the same way as that of radiation.
We still call it as the radiation dominated epoch for simplicity. 

Although the scalar field component is not negligible at early time, the model is not one of the early dark energy models because $w_{\phi}$ is not less than $-1/3$ in the deep radiation dominated era. By the same reason only the fixed point (d) can be the matter-dominated epoch (or rigorously $\phi$-matter dominated epoch). The fixed point (b) cannot be the dark energy dominated epoch because $w_{\phi}$ is not less than $-1/3$. The fixed point (e) requires $\lambda < 8$ to provide the accelerated expansion. However, it is contradictory to the constraint on the fixed point (c). We adopt the Big Bang Nucleosysthesis (BBN) constraint on the quintessence model, $\Omega_{\phi} \vert_{\rm BBN} < 0.045$ \cite{Bean:2001wt}, then the scalar field density parameter of the fixed point (c), $\Omega_{\phi} = 10/\lambda$, leads to
\bea
\lambda > 222.22 \,. \label{BBN}
\ena
Therefore the dark energy dominated epoch corresponds only to the fixed point (a) which is the de Sitter fixed point because $\Omega_{\phi} = 1$ and $w_{\phi} = -1$.

\subsubsection{Inverse power-law potential}
 
Considering the inverse power-law potential as the following form
\bea
V(\phi) = \frac{M^{4+n}}{\phi^n} \,,
\ena
where $M$ is a constant with dimension of mass, and $n > 0$. We find $\Gamma = (n+1)/n$ or $(\Gamma - 1) = 1/n$. In this case $\lambda$ is not a constant, we then need to solve the Eq. (\ref{auto5}) along with the previous $4$ autonomous equations, (\ref{auto1}) - (\ref{auto4}). Setting $dx_1/dN = dx_2/dN = dx_3/dN = dx_4/dN = d\lambda/dN= 0$, we find three fixed points as shown in Table \ref{fixedpoints2}.


\begin{table}[h]
\begin{center}
  \begin{tabular}{c c c c c c c c c c c}
\hline 
\hline 
& & & & & & & & & & \\[-.5em]
Fixed point& $x_1$ & $x_2$ & $x_3$ & $x_4$ & $\lambda$ & $\Omega_m$ & $\Omega_r$ & $\Omega_{\phi}$ & $w_{\phi}$ & $w_{\rm eff}$
\\[.5em]
\hline
& & & & & & & & & & \\[-.5em]
(f) & $1+x_3$ & $3x_3$ &  & $0$ & $0$ & $0$ & $0$ & $1$ & $-1$ & $-1$
\\[.5em]
& & & & & & & & & & \\[-.5em]
(g) & $0$ & $-3$ & $-1$ & $0$ & $0$ & $0$ & $0$ & $1$ & $-1$ & $-1$
\\[.5em]
& & & & & & & & & & \\[-.5em]
(h) & $0$ & $0$ & $-1$ & $0$ & $0$ & $0$ & $0$ & $1$ & $1$ & $1$
\\[.5em]
\hline 
 \end{tabular}
    \caption{The fixed points, the density parameters, and the equation of state parameters of the cuscuta-galileon with the inverse power-law potential for any integer $n$.}
\label{fixedpoints2}
\end{center}
\end{table}


The fixed points (f) and (g) are possible to be the dark energy dominated epoch 
just as a conventional quintessence model.
 The point (h) describes the stiff-matter universe and then 
does not match with any thermal history of the Universe. Since the autonomous system of the inverse power-law potential does not provide the radiation dominated and matter dominated eras, we will no longer consider this case for the rest of this paper.

In the next subsection we will check stability of the fixed points of the cuscuta-galileon with the exponential potential.

\subsection{Stability of Fixed Points}

In order to discuss the roles of the above fixed points in the history of the universe, we 
have to discuss the stability of the fixed points.  
We then perturb the variables around fixed points. 


The $4$ autonomous equations (\ref{auto1})-(\ref{auto4}) with $4$ parameters are described as
\beann
\frac{dx_1}{dN} &=& \mathcal{A} (x_1, x_2, x_3, x_4) \,, \\
\frac{dx_2}{dN} &=& \mathcal{B} (x_1, x_2, x_3, x_4) \,, \\
\frac{dx_3}{dN} &=& \mathcal{C} (x_1, x_2, x_3, x_4) \,, \\
\frac{dx_4}{dN} &=& \mathcal{D} (x_1, x_2, x_3, x_4) \,.
\enann
Considering the linear perturbation around the fixed points, $x \rightarrow x^{\rm (FP)} + \delta x$, we obtain the first order coupled differential equations:
\bea
\frac{d}{dN} \begin{pmatrix} \delta x_1 \\ \delta x_2 \\ \delta x_3 \\ \delta x_4 \end{pmatrix} =
 \mathcal{M} \begin{pmatrix} \delta x_1 \\ \delta x_2 \\ \delta x_3 \\ \delta x_4 \end{pmatrix} \,, \label{firstordercouple}
\ena
where the matrix $\mathcal{M}$ depends on the fixed points as 
\beann
\mathcal{M} = \left.\begin{pmatrix} \frac{\partial \mathcal{A}}{\partial x_1} && \frac{\partial \mathcal{A}}{\partial x_2} &&  \frac{\partial \mathcal{A}}{\partial x_3} &&  \frac{\partial \mathcal{A}}{\partial x_4} \\[.5em]
\frac{\partial \mathcal{B}}{\partial x_1} && \frac{\partial \mathcal{B}}{\partial x_2} &&  \frac{\partial \mathcal{B}}{\partial x_3} &&  \frac{\partial \mathcal{B}}{\partial x_4} \\[.5em]
\frac{\partial \mathcal{C}}{\partial x_1} && \frac{\partial \mathcal{C}}{\partial x_2} &&  \frac{\partial \mathcal{C}}{\partial x_3}  &&  \frac{\partial \mathcal{C}}{\partial x_4} \\[.5em]
\frac{\partial \mathcal{D}}{\partial x_1} && \frac{\partial \mathcal{D}}{\partial x_2} &&  \frac{\partial \mathcal{D}}{\partial x_3}  &&  \frac{\partial \mathcal{D}}{\partial x_4}
\end{pmatrix}\right|_{x_1^{\rm (FP)},x_2^{\rm (FP)},x_3^{\rm (FP)},x_4^{\rm (FP)}} \,.
\enann
The eigen functions of the Eq. (\ref{firstordercouple}) are given by
\bea
\delta x_i^{\,(a)} \propto e^{\mu^{(a)} N} \,,~~(a=1,\cdots 4)
\ena
where $\mu^{(a)}$ are the eigenvalues of the matrix $\mathcal{M}$.

 If all eigenvalues are negative, we find a stable fixed point. In the case of complex eigenvalues, if all real parts are negative, the fixed point is a stable spiral point, whereas if all of them are positive, the fixed point is an unstable point or unstable spiral point for complex eigenvalues. 
 If at least one eigenvalue but not all  is positive (or gives a positive real part), 
 the fixed point is a saddle point.

We summarize all eigenvalues of the fixed points in the Table \ref{fixedpoints}:
\beann
{\rm (a)} &:& \Big(-4,-3,-3, - \frac{\lambda}{2}\Big) \,, \\
{\rm (b)} &:& \Big(~3,~3,~2,~6 - \frac{\lambda}{2}\Big) \,, \\
{\rm (c)} &:& \Big(~2,~1, - \frac{1}{2} \pm \frac{\sqrt{41-4\lambda}}{2}\Big) \,, \\
{\rm (d)} &:& \Big(~\frac{3}{2},-1, - \frac{3}{4} \pm \frac{\sqrt{3(75 - 8\lambda)}}{4} \Big) \,, \\
{\rm (e)} &:& \Big(~\lambda-10,~\frac{\lambda}{2} - 6,~\lambda - 9,~\frac{\lambda}{2} - 3 \Big) \,.
\enann
Consequently, the fixed point (a) is a stable fixed point, the fixed point (b), (c), and (d) are saddle points, and the fixed point (e) is an unstable point. 
Remind that the fixed point (b) does not relate to any thermal history of the Universe, and the fixed point (e) requires $\lambda < 8$ to give the accelerated expansion. 
However, since we need the fixed point (c) to be the radiation dominated epoch, it must satisfy the condition, $\lambda > 222.22$, from the BBN constraint. 
Therefore, if we start from the fixed point (c), the cosmological sequence is 
\beann
{\rm (c)} \rightarrow {\rm (d)} \rightarrow {\rm (a)} \,.
\enann

\section{Degrees of Freedom} 
\label{dof}

 As shown in Appendix \ref{independenteq}, 
we find the dynamical equation for the scalar field 
in a homogeneous field in FLRW universe.
Hence, first we have to check the degree of freedom for the present model.

In this section we will use the Hamiltonian formalism to find degrees of freedom. 
According to the Refs. \cite{Tsujikawa:2014mba,Kase:2014yya,Kase:2014cwa} the action (\ref{action}) can be written in the Arnowitt-Deser-Misner (ADM) form as
\bea
S &=& \int dt d^3 x N \sqrt{h} \Big[\frac{1}{2}\mpl^2 \left({^3 R} + K_{ij}K^{ij} - K^2 \right) \nn
& & + \left(\frac{a_2 \lvert \dot\phi \rvert}{N} -V(\phi) \right) + \left(- \frac{2 a_3 \lvert \dot\phi \rvert}{N} + C \right) K \Big] \,, \nn
\ena
where $^3 R$ is the three-dimensional Ricci scalar, $K_{ij}$ is the extrinsic curvature, $K \equiv K_{ij} h^{ij}$, and $C$ is a constant. Note that in this section we will not consider contribution from the matter field. 

In the ADM Language the fundamental variables are $N$, $N^i$, and $h^{ij}$ where they are the lapse function, the shift vector, and the three-dimensional metric, respectively. Following calculations in Ref. \cite{Lin:2014jga} we choose the unitary gauge, $\phi = \phi(t)$, then the scalar field is merely time. Hence we have only 10 fundamental variables, whose conjugate momenta are
\bea
\pi_N &=& \frac{\partial \cal{L}}{\partial \dot N} = 0 \,, ~~~ \pi_i = \frac{\partial \cal{L}}{\partial \dot N^i} = 0 \,, \\
\pi^{ij} &=& \frac{\partial \cal{L}}{\partial \dot h_{ij}} 
= \frac{1}{2}\mpl^2 \sqrt{h} \left(K^{ij} - h^{ij} K \right) \nn 
& & ~~~~~~~~~~ + \frac{1}{2} \sqrt{h} \left( - \frac{2 a_3 \lvert \dot\phi \rvert}{N} + C\right) h^{ij} \,.
\ena
Using the Legendre transformation, the Hamiltonian of the cuscuta-galileon gravity is given by
\bea
H = \int d^3 x \left({\cal H} + N^i {\cal H}_i + \lambda_N \pi_N + \lambda^i \pi_i \right) \,,
\ena
where $\lambda_N$ and $\lambda^i$ are Lagrange multipliers, and 
\bea
{\cal H} &=& N \sqrt{h} \left[\frac{2}{\mpl^2} \left(\frac{\pi^{ij} \pi_{ij}}{h} - \frac{\pi^2}{2h}\right) - \frac{1}{2} \mpl^2 {}^3 R \right. \nn
& & \left. - \left(\frac{a_2 \lvert \dot\phi \rvert}{N} -V(\phi) \right) + \frac{\pi}{\mpl^2 \sqrt{h}} \left(- \frac{2 a_3 \lvert \dot\phi \rvert}{N} + C \right) \right. \nn 
& & \left. - \frac{3}{4\mpl^2} \left(- \frac{2 a_3 \lvert \dot\phi \rvert}{N} + C \right)^2 \right] \,, \\
{\cal H}_i &=& - 2 h_{ik} D_j \pi^{kj} \,.
\ena
$\pi$ is a trace of the $\pi_{ij}$, and $D_j$ is the three-dimensional covariant derivative. Although the form of ${\cal H}_i$ is the same as in GR, it is not a first-class constraint because the ${\cal H}$ is not a linear function of $N$ (see Ref. \cite{Lin:2014jga}). In order to obtain ${\cal H}_i$ as the first-class constraint we need to add additional terms which vanish weakly to the Hamiltonian as
\bea
{\cal \bar H}_i \equiv {\cal H}_i + \pi_N \partial_i N \,.
\ena 
Using conservation of the $4$ primary constraints, $\pi_N = 0, \pi_i = 0$, we find secondary constraints as
\bea
0 &=& \frac{d\pi_N}{dt} = - \frac{\partial H}{\partial N} \approx - \frac{\partial {\cal H}}{\partial N} \equiv {\cal C} ~\rightarrow~ {\cal C} \approx 0 \,, \\
0 &=& \frac{d\pi_i}{dt} = - \frac{\partial H}{\partial N^i} \approx {\cal \bar H}_i ~\rightarrow~ {\cal \bar H}_i \approx 0 \,.
\ena
The notation $\approx$ means the weak equality, i.e. it is the equality on the constraint surface in phase space. 

We can check whether these constraints are first-class or second-class by using the Poisson bracket which is given by
\bea
\{F,G\} \equiv \int d^3 y \Big[\frac{\delta F}{\delta N(y)} \frac{\delta G}{\delta \pi_N (y)} &-& \frac{\delta F}{\delta \pi_N (y)} \frac{\delta G}{\delta N (y)} \nn
+ \frac{\delta F}{\delta N^i(y)} \frac{\delta G}{\delta \pi_i (y)} &-& \frac{\delta F}{\delta \pi_i (y)} \frac{\delta G}{\delta N^i (y)} \nn
+ \frac{\delta F}{\delta h_{ij} (y)} \frac{\delta G}{\delta \pi^{ij} (y)} &-& \frac{\delta F}{\delta \pi^{ij} (y)} \frac{\delta G}{\delta h_{ij} (y)} \Big] \,. \nn
\ena
Therefore, we find
\bea
\{\pi_i (x), \pi_N (x^{\prime})\} &=& 0 \,, \\
\{\pi_i (x), {\cal \bar H}_j (x^{\prime})\} &=& 0 \,, \\
\{\pi_i (x), {\cal C} (x^{\prime})\} &=& 0 \,, \\
\{ {\cal \bar H}_i [f^i], \bar \pi_N [\varphi] \} &=& \int d^3 y \pi_N f^i \partial_i \varphi \approx 0 \,, \\
\{ {\cal \bar H}_i [f^i], {\cal C} [\varphi] \} &=& \int d^3 y {\cal C} f^i \partial_i \varphi \approx 0 \,, \\
\{\pi_N (x), {\cal C} (x^{\prime})\} &=& \frac{\partial^2 {\cal H}}{\partial N^2} \delta(x - x^{\prime}) \,.
\ena
Some Poisson brackets we use the smeared constraint form defined as
\bea
{\cal \bar H}_i [f^i] &\equiv& \int d^3 x f^i (x) {\cal \bar H}_i (x) \,, \\
\bar \pi_N [\varphi] &\equiv& \int d^3 x \varphi (x) \pi_N (x) \,, \\
{\cal C} [\varphi] &\equiv& \int d^3 x \varphi (x) {\cal C} (x) \,.
\ena
The Poisson brackets are vanished except the last one because the ${\cal H}$ of the cuscuta-galileon model gives 
\bea
\frac{\partial^2 {\cal H}}{\partial N^2} \not= 0 \,.
\ena
As a result, we have $10$ variables which correspond to $20$ dimensions in phase space with $8$ constraints where $\pi_i$, ${\cal \bar H}_i$ are the first-class constraints, and $\pi_N$, ${\cal C}$ are the second-class constraints. Consequently, the number degrees of freedom of the cuscuta-galileon gravity is given by
\bea
{\rm d.o.f.} &=& \frac{1}{2} ({\rm variables} \times 2 - {\rm 1st~class} \times 2 - {\rm 2nd~class}) \nn
&=& \frac{1}{2} (20 - 6 \times 2 - 2)  \nn
&=& 3 \,.
\ena

We then find that  the cuscuta-galileon gravity has three d.o.f. 
instead of two.
As a result, the present model is neither included in 
MMG nor  a subclass of the extended cuscuton gravity, 
but it is rather in a subclass of the Horndeski theories.

\section{Cosmic Evolution in the cuscuta-galileon theory}
\label{cosmic evolution}
\subsection{Numerical Solution}
\label{numer}

\begin{figure}[h]
	\centerline{\includegraphics[width=7cm]{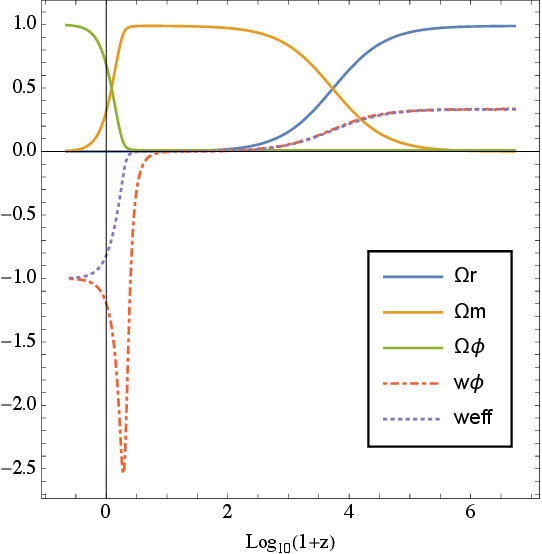}}
	\caption{The evolution of the density parameters and the equation of state parameters where we set $x_1 - 0.002 = 1 \times 10^{-6}$, $x_2 = - 1 \times 10^{-13}$, $x_3 + 0.008 = 1 \times 10^{-6}$, and $x_4 = 0.989$ at $\log_{10} (1+z) = 6.73$.}
\label{evolution}
\end{figure}

In this section, 
solving the autonomous equations (\ref{auto1}) - (\ref{auto4}) of the cuscuta-galileon with the exponential potential numerically, we 
discuss how the Universe evolves in the present model. 
We set $\lambda = 10^3$ and choose initial conditions near the fixed point (c) (i.e., starting from the radiation dominated epoch). The evolution of the density parameters and the equation of state parameters according to the Eqs. (\ref{omegam}), (\ref{omegarandphi}), (\ref{weff}), and (\ref{wphi}) are shown as Fig. \ref{evolution}.

\begin{figure}[h]
	\centerline{\includegraphics[width=7cm]{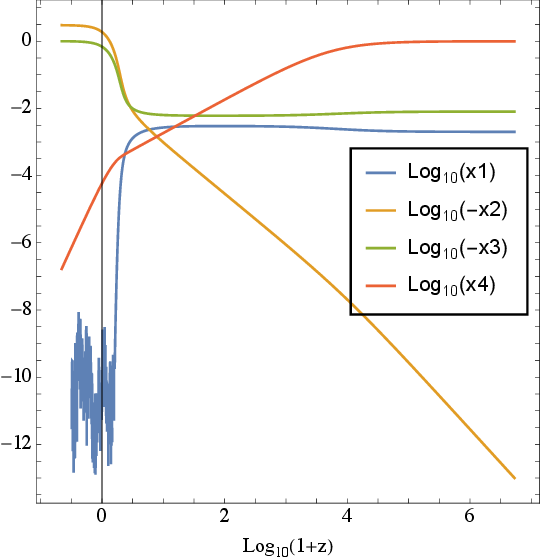}}
	\caption{The evolution of the dynamical parameters where the initial conditions are the same as the Fig. \ref{evolution}.}
\label{logx}
\end{figure}

The Fig. \ref{evolution} reveals that the evolution of the cuscuta-galileon with the exponential potential corresponds to the thermal history of the Universe correctly. This result is consistent with the stability analysis on the fixed points. The $w_{\phi} = 1/3$ in the radiation dominated era, and then it is around zero in the matter dominated era. However, before approaching the de Sitter fixed point, the $w_{\phi}$ crosses the cosmological constant boundary, $w_{\Lambda} = -1$, and then approaching $-1$ at late time. The large negative value can be understood by considering the evolution plot of the dynamical parameters as Fig. \ref{logx}.

According to the Fig. \ref{logx}, the $x_1$ and the $x_4$ tend to zero around the end of the matter dominated epoch, then the Eq. (\ref{wphi}) becomes $w_{\phi} \simeq - w_{\rm eff} / x_3$. Since $|x_3| < 1$, we find $|w_{\phi}| > |w_{\rm eff}|$, and at late time the $|x_3| \rightarrow 1$, thus $|w_{\phi}| \simeq |w_{\rm eff}|$. Therefore, we obtain a large negative value of the $w_{\phi}$ around the end of the matter dominated era, and then it approaches to the $w_{\rm eff}$ at late time.

Note that the fine-tuning parameters are the amount of the radiation component, $x_4$, and the ratio of velocity of the scalar field and the Hubble parameter squared, $x_2$, in the radiation dominant to have long enough the matter dominant epoch. If $x_4$ is larger or $x_2$ is more negative, the matter dominant era will be shorter. It is then inconsistent with observations that the age of matter-radiation equality is around $z \approx 3300$. The other parameters are more flexible, for example, $x_1$ and $x_3$ can be around $10^{-5}$ from the fixed point (c), we still obtain the sequence of the cosmic evolution properly. However, there is a small oscillations on the value of $w_{\phi}$ in this case. 

\subsection{Ghosts and Laplacian instability} 
\label{perturbations}

As shown in  \S. \ref{dof}, 
 this model includes  an additional degrees of freedom (a scalar field) 
 in the present model.
Then we have to check whether there exists no ghost or Laplacian instability 
in our cosmic evolution.
Since there are three degrees of freedom, we expect that there exist 
scalar perturbations as well as tensor perturbations.

According to Refs. \cite{DeFelice:2011bh,Gergely:2014rna,Tsujikawa:2014mba,Kase:2014yya,Kase:2014cwa,Gleyzes:2013ooa} the second order action of the tensor perturbations is 
\bea
S_2^T = \int d^4 x \frac{a^3}{4} L_S \Big[ \dot \gamma_{ij}^2 - c_T^2 \frac{(\partial_k \gamma_{ij})^2}{a^2} \Big] \,,
\ena
where $\gamma_{ij}$ is the tensor perturbations which satisfies transverse and traceless conditions, $L_S$ relates to the action in the background level, and $c_T^2$ is a sound speed squared in the tensor mode which also relates to the action in the background level. In order to avoid ghosts the coefficient in front of the term $\dot \gamma_{ij}^2$ must be positive, thus we need the $L_S > 0$. Similarly, we require the $c_T^2 > 0$ to avoid the Laplacian instability.

For simplicity we use notations as the Ref. \cite{DeFelice:2011bh} where they correspond to the action of the cuscuta-galileon (\ref{action}) as follows
\bea
{\cal W}_1 &=& \mpl^2 \,, \\
{\cal W}_2 &=& 2 \mpl^2 H + 2 \dot \phi a_3 = (2 + x_3) \mpl^2 H \,, \\
{\cal W}_3 &=& -9 \mpl^2 H^2 - 18 H \dot\phi a_3 \nn 
&=& (-9 - 9 x_3) \mpl^2 H^2 \,, \\
{\cal W}_4 &=& \mpl^2 \,,
\ena
where $L_S \equiv {\cal W}_1 /2$ and $c_T^2 \equiv {\cal W}_4 / {\cal W}_1$. We then find 
\bea
L_S = \frac{1}{2}\mpl^2 > 0 \,, ~~~ c_T^2 = 1 \,.
\ena
Consequently, the cuscuta-galileon gravity satisfies the no ghosts and no Laplacian instability conditions of the tensor mode.

For the scalar perturbations it is similar to the tensor perturbations. In order to avoid the ghosts we require 
\bea
Q_S \equiv \frac{{\cal W}_1 (4{\cal W}_1 {\cal W}_3 +9 {\cal W}_2^2)}{3{\cal W}_2^2} = \frac{3x_3^2}{(2 + x_3)^2} > 0 \,, \label{qs}
\ena
and the sound speed squared in the scalar mode must be greater than zero to avoid the Laplacian instability:
\begin{widetext}
\bea
c_S^2 &\equiv& \frac{3\Big(- 2{\cal W}_1^2 \dot {\cal W}_2 + 2 {\cal W}_1^2 {\cal W}_2 H - {\cal W}_2^2 {\cal W}_4 - 2{\cal W}_1^2 \Big((1+w_m)\rho_m + (1+w_r)\rho_r\Big)\Big)}{{\cal W}_1 (4{\cal W}_1 {\cal W}_3 +9 {\cal W}_2^2)} 
= \frac{2 x_2 - x_3 (8 + x_3)}{3x_3^2} > 0 \,. \label{cssq}
\ena
\end{widetext}
Other conditions involving the existence of nonrelativistic matter and radiation fluids are automatically satisfied when we choose forms of the k-essence type perfect fluid as the Ref. \cite{Kase:2014yya}. 

\begin{figure}[h]
	\centerline{\includegraphics[width=7cm]{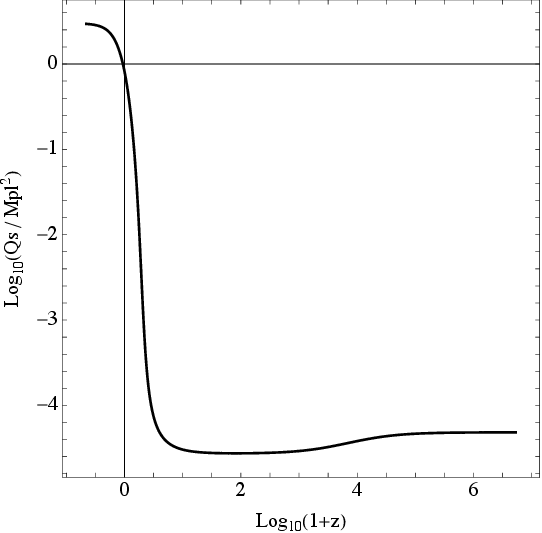}}
	\centerline{\includegraphics[width=7cm]{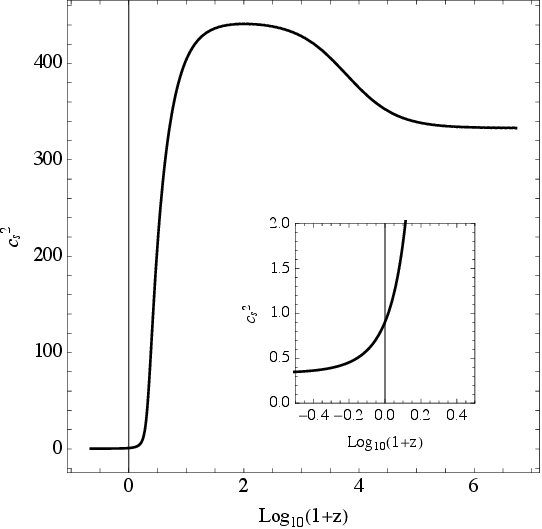}}
	\caption{The evolutions of the $Q_S$ as Eq. (\ref{qs}) in log scale and the $c_S^2$ as Eq. (\ref{cssq}) where the initial conditions are the same as the Fig. \ref{evolution}. }
\label{qscsplot}
\end{figure}

The Fig. \ref{qscsplot} reveals that the cuscuta-galileon gravity has no ghosts and the Laplacian instability in the scalar mode. In matter dominant and radiation dominant the sound speed squared is greater than unity because $|x_3| \ll 1$ in the denominator of the Eq. (\ref{cssq}), while  $x_2$ is about zero. 

The results of the scalar perturbations reveal that there is a scalar degree of freedom propagating in this model,
 which is consistent with the analysis in the previous section.

We can construct a viable cosmological model in the cuscuta-galileon gravity theory.
We show the cosmological evolution from radiation dominated era to de Sitter expansion stage  via matter dominated era.

\subsection{Observational constraints} 
\label{observational_constraints}

\begin{figure}[h]
	\centerline{\includegraphics[width=7cm]{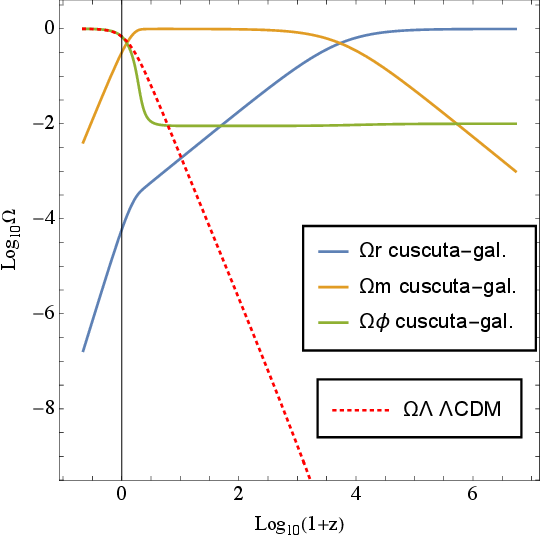}}
	\caption{The evolution of the density parameters in log scale where the initial conditions are the same as the Fig. \ref{evolution}. }
\label{logevolution}
\end{figure}

However if we look into the detail, we find
 that the cuscuta-galileon gravity may not satisfy observations by several reasons. First, the Planck $2018$ results \cite{Aghanim:2018eyx} reveal that the dark energy equation of state parameter is $w_{\rm DE} = -1.028 \pm 0.031$, it is consistent with the cosmological constant, while the cuscuta-galileon gives $w_{\phi} = -1.196$ at $\Omega_m = 0.315$. Second, there is a large amount of the dark energy component comparing to the cosmological constant in the matter and radiation dominated eras as Fig. \ref{logevolution}. 

From the Lambda-Cold Dark Matter ($\Lambda$CDM) model, the density parameter of the cosmological constant is given by
\bea
\Omega_{\Lambda} (z) = \frac{\Omega_{\Lambda}^{(0)}}{\Omega_m^{(0)}(1+z)^3 + \Omega_r^{(0)}(1+z)^4 + \Omega_{\Lambda}^{(0)}} \,,
\ena
where $\Omega_m^{(0)} = 0.315$, $\Omega_r^{(0)} = 9 \times 10^{-5}$, and $\Omega_{\Lambda} = 0.685$ according to the Planck $2018$ results. In the Fig. \ref{logevolution} we find that the $\Omega_{\phi} \sim \mathcal{O} (10^{-2})$ in the matter and radiation dominated epochs, whereas $\Omega_{\Lambda}$ is utterly small, for instance, at the last scattering surface, $z \approx 1090$, the $\Omega_{\Lambda} \sim \mathcal{O} (10^{-9})$. It is obvious that the model is different from the $\Lambda$CDM model. Therefore, the cuscuta-galileon gravity is likely not to satisfy the observations which prefer the $\Lambda$CDM model, such as the Cosmic Microwave Background (CMB) observations. Although rigorous calculations and global fitting with observational data are required, they are beyond the scope of this paper. Lastly, if we increase the $\lambda$ in order to obtain the lower $\Omega_{\phi}$, such as $\lambda = 10^4$, we find $\Omega_{\phi} \sim \mathcal{O} (10^{-3})$, the equation of state parameter of the scalar field will be more negative and more deviate from the observational value because $|x_3| \sim \mathcal{O} (10^{-3}) \ll 1$ around the end of the matter dominated epoch.


\section{Conclusions} 
\label{conclusions}

In this work we study cosmological dynamics of the cuscuta-galileon gravity. The model was proposed in the Ref. \cite{deRham:2016ged} as galileon generalization of the cuscuton model which is free from the caustic singularities in flat space-time. In the case without a potential term the equation of motion of the cuscuta-galileon does not depend on a scalar field,
finding that there exists only the de Sitter expansion under the flat FLRW background. 
Thus in order to obtain the radiation and matter dominated eras we need to add a potential term, for which
 we consider two cases: an exponential potential and an inverse power-law  potential.
  Using the dynamical system approach and studying stability of fixed points of the autonomous system,
   we find that only the exponential potential case can provide 
  a proper sequence of the thermal history of the Universe successfully. 
  
Even though there is no second-order time derivatives in the equation of motion, the results of the scalar perturbation reveal that there is a scalar degree of freedom propagating in this model. This is confirmed by using the Hamiltonian analysis where we find that the 
cuscuta-galileon gravity actually has three degrees of freedom
and belongs to a subclass of Horndeski theories. 
In order to discuss the similar cuscuta-galileon theory with only two d.o.f., we have to include an additional kinetic term, which was discussed in \cite{Maeda:2022ozc}.

In the perturbation level,
the conditions for avoidance of ghosts and the Laplacian instability in the tensor mode are automatically satisfied by the form of the action. In the scalar mode we find that there is no ghost and  Laplacian instabilities in the present cosmological model. 

  However, the detail numerical analysis reveals that there appears a large amount of the dark energy component in the matter and radiation dominated eras comparing 
  to that in  the $\Lambda$CDM model. Therefore, the present cuscuta-galileon gravity may not satisfy the observational constraints.

\section*{Acknowledgements}

S.P. would like to thank Aya Iyonaga and Shinji Tsujikawa for useful discussions.
This work was supported in part by a Waseda University Grant for Special Research Project (No. 2020C-774)
and by JSPS KAKENHI Grant Numbers JP17H06359 and JP19K03857.

\appendix

\section{Cuscuta-galileon gravity without a potential term} 
\label{cuscutawithoutpotential}

Considering the action of the cuscuta-galileon gravity as Ref. \cite{deRham:2016ged} in curved space-time up to cubic order:
\bea
S &=& \int d^4 x \sqrt{-g} \Big[ \frac{1}{2} \mpl^2 R + a_2 \sqrt{-X} + a_3 \ln \Big(-\frac{X}{\Lambda^4}\Big) \square \phi \Big] \nn 
& & + S_M (g_{\mu\nu}, \psi_M) \,,
\ena
Substituting the flat FLRW metric, $ds^2 = -N(t)^2 dt^2 + a(t)^2 \delta_{ij} dx^i dx^j$, into the above action, and choosing the unitary gauge, $\phi = \phi(t)$. Varying the action with respect to $\phi$, after setting $N = 1$ the equation of motion of the scalar field is given by
\bea
6a_3 H^2 + 2a_3 \dot H - a_2 H {\rm sgn} (\dot\phi) = 0 \,. 
\ena
Since the above equation depends on $H$ only, we can integrate it directly. The evolution of the Hubble parameter is
\bea
H(t) = \frac{a_2 {\rm sgn} (\dot\phi)}{6a_3 - e^{-(a_2 {\rm sgn} (\dot\phi) (t + C)/ 2a_3)}} \,,
\ena
where $C$ is a constant of integration. If $a_2 {\rm sgn} (\dot\phi)/ 2a_3 > 0$, we find 
\bea
\lim_{t \rightarrow \infty} H(t) = \frac{a_2 {\rm sgn} (\dot\phi)}{6a_3} = {\rm constant} \,. \label{analyticsolution} 
\ena
Then, we obtain the de Sitter solution at late time. If $a_2 {\rm sgn} (\dot\phi)/ 2a_3 < 0$, we find
\bea
\lim_{t \rightarrow \infty} H(t) = 0 \,. 
\ena
This is the static universe solution; however, this solution is contradict with observations. We thus accept only the de Sitter solution.

Using the dynamical system approach as the section \ref{dynamical}, 
if the cuscuta-galileon model does not have a potential term, then the $x_1 = 0$ and the $\lambda$ is undefined.
Therefore we have only $3$ autonomous equations (\ref{auto2}) - (\ref{auto4}) with $3$ parameters. Setting $dx_2/dN = dx_3/dN = dx_4/dN = 0$, we find fixed points as Table \ref{fixedpoints3}.

\begin{table}[h]
\begin{center}
  \begin{tabular}{c c c c c c c c c}
\hline 
\hline 
& & & & & & & & \\[-.5em]
Fixed point& $x_2$ & $x_3$ & $x_4$ & $\Omega_m$ & $\Omega_r$ & $\Omega_{\phi}$ & $w_{\phi}$ & $w_{\rm eff}$
\\[.5em]
\hline
& & & & & & & & \\[-.5em]
(i) & $-3$ & $-1$ & $0$ & $0$ & $0$ & $1$ & $-1$ & $-1$
\\[.5em]
& & & & & & & & \\[-.5em]
(j) & $0$ & $-1$ & $0$ & $0$ & $0$ & $1$ & $1$ & $1$
\\[.5em]
\hline 
 \end{tabular}
    \caption{The fixed points, the density parameters, and the equation of state parameters of the cuscuta-galileon without a potential.}
\label{fixedpoints3}
\end{center}
\end{table}

The fixed point (i) can be the dark energy dominated epoch, whereas the point (j) does not match with any thermal history of the Universe. Then we obtain only the de Sitter expansion in the cuscuta-galileon gravity without a potential term. This result is consistent with the analytic solution (\ref{analyticsolution}).

\section{Two independent equations}
\label{independenteq}

There are three basic equations, but they are not independent.
For example, taking the time derivative of Eq.(\ref{1stFriedmann}) and eliminating $
\ddot{\phi}$ by use of Eq. (\ref{2ndFriedmann}) and the equations of $\dot \rho_i$ ($i=m, r$),
i.e., $$\dot\rho_i+3H(\rho_i+P_i)=0\,,$$ we obtain
Eq. (\ref{EOM}). 

In fact we obtain the following two independent equations:

\begin{widetext}
\bea
&&
\ddot\phi-{3a_3\over M_{PL}^2}\dot\phi^2
+\left(3\dot \phi+{a_2\over 2a_3^2}M_{PL}^2 {\rm sgn} (\dot\phi) \right)\sqrt{\left({a_3\over M_{PL}^2}\dot\phi\right)^2+{1\over 3M_{PL}^2}\left(\rho_m+\rho_r+V\right)}\nonumber \\
&&~~~
-{1\over 2a_3}\left(\rho_m-P_m+\rho_r-P_r+2V\right)+{M_{PL}^2\over 6a_3^2}V_{,\phi}=0\,,
\label{eq:phi}
\\
&&H\equiv {\dot a\over a}=
-{a_3\over M_{PL}^2}\dot\phi+\sqrt{\left({a_3\over M_{PL}^2}\dot\phi\right)^2+{1\over 3M_{PL}^2}\left(\rho_m+\rho_r+V\right)}\,.
\nonumber \\
&&~
\label{eq:Hubble}
\ena

\end{widetext}
Eq. (\ref{eq:phi}) is the second order differential equation for $\phi$, while 
Eq. (\ref{eq:Hubble}) is the first differential equation for $a$.
$P_m\,,\rho_m$ and $P_r\,,\rho_r$ are given by a scale factor $a$ as
\bea
P_m=0 \,, ~\rho_m\propto a^{-3} \,,
\ena
and 
\bea
P_r={\rho_r\over 3} \,, ~\rho_r\propto a^{-4}
\,.
\ena
Hence once we know the initial values of $\phi\,,\dot\phi\,,\rho_m\,,\rho_r$ and $a$,
we find the time evolution of those variables.



\begin{thebibliography}{99}

\bibitem{Riess:1998cb}
A.~G.~Riess \textit{et al.} [Supernova Search Team],
Astron. J. \textbf{116} (1998), 1009-1038
doi:10.1086/300499
[arXiv:astro-ph/9805201 [astro-ph]].

\bibitem{Perlmutter:1998np}
S.~Perlmutter \textit{et al.} [Supernova Cosmology Project],
Astrophys. J. \textbf{517} (1999), 565-586
doi:10.1086/307221
[arXiv:astro-ph/9812133 [astro-ph]].

\bibitem{Horndeski:1974wa}
G.~W.~Horndeski,
Int. J. Theor. Phys. \textbf{10} (1974), 363-384
doi:10.1007/BF01807638

\bibitem{Deffayet:2011gz}
C.~Deffayet, X.~Gao, D.~A.~Steer and G.~Zahariade,
Phys. Rev. D \textbf{84} (2011), 064039
doi:10.1103/PhysRevD.84.064039
[arXiv:1103.3260 [hep-th]].

\bibitem{Kobayashi:2011nu}
T.~Kobayashi, M.~Yamaguchi and J.~Yokoyama,
Prog. Theor. Phys. \textbf{126} (2011), 511-529
doi:10.1143/PTP.126.511
[arXiv:1105.5723 [hep-th]].

\bibitem{Heisenberg:2014rta}
L.~Heisenberg,
JCAP \textbf{05} (2014), 015
doi:10.1088/1475-7516/2014/05/015
[arXiv:1402.7026 [hep-th]].

\bibitem{deRham:2010ik}
C.~de Rham and G.~Gabadadze,
Phys. Rev. D \textbf{82} (2010), 044020
doi:10.1103/PhysRevD.82.044020
[arXiv:1007.0443 [hep-th]].

\bibitem{deRham:2010kj}
C.~de Rham, G.~Gabadadze and A.~J.~Tolley,
Phys. Rev. Lett. \textbf{106} (2011), 231101
doi:10.1103/PhysRevLett.106.231101
[arXiv:1011.1232 [hep-th]].

\bibitem{Will:2014kxa}
C.~M.~Will,
Living Rev. Rel. \textbf{17} (2014), 4
doi:10.12942/lrr-2014-4
[arXiv:1403.7377 [gr-qc]].

\bibitem{Khoury:2003aq}
J.~Khoury and A.~Weltman,
Phys. Rev. Lett. \textbf{93} (2004), 171104
doi:10.1103/PhysRevLett.93.171104
[arXiv:astro-ph/0309300 [astro-ph]].

\bibitem{Khoury:2003rn}
J.~Khoury and A.~Weltman,
Phys. Rev. D \textbf{69} (2004), 044026
doi:10.1103/PhysRevD.69.044026
[arXiv:astro-ph/0309411 [astro-ph]].

\bibitem{Hinterbichler:2010es}
K.~Hinterbichler and J.~Khoury,
Phys. Rev. Lett. \textbf{104} (2010), 231301
doi:10.1103/PhysRevLett.104.231301
[arXiv:1001.4525 [hep-th]].

\bibitem{Brax:2010gi}
P.~Brax, C.~van de Bruck, A.~C.~Davis and D.~Shaw,
Phys. Rev. D \textbf{82} (2010), 063519
doi:10.1103/PhysRevD.82.063519
[arXiv:1005.3735 [astro-ph.CO]].

\bibitem{Burrage:2014uwa}
C.~Burrage and J.~Khoury,
Phys. Rev. D \textbf{90} (2014) no.2, 024001
doi:10.1103/PhysRevD.90.024001
[arXiv:1403.6120 [hep-th]].

\bibitem{Panpanich:2019rij}
S.~Panpanich, S.~Ponglertsakul and K.~Maeda,
Phys. Rev. D \textbf{100} (2019) no.4, 044038
doi:10.1103/PhysRevD.100.044038
[arXiv:1902.00265 [gr-qc]].

\bibitem{Brax:2012jr}
P.~Brax, C.~Burrage and A.~C.~Davis,
JCAP \textbf{01} (2013), 020
doi:10.1088/1475-7516/2013/01/020
[arXiv:1209.1293 [hep-th]].

\bibitem{Babichev:2009ee}
E.~Babichev, C.~Deffayet and R.~Ziour,
Int. J. Mod. Phys. D \textbf{18} (2009), 2147-2154
doi:10.1142/S0218271809016107
[arXiv:0905.2943 [hep-th]].

\bibitem{Babichev:2013usa}
E.~Babichev and C.~Deffayet,
Class. Quant. Grav. \textbf{30} (2013), 184001
doi:10.1088/0264-9381/30/18/184001
[arXiv:1304.7240 [gr-qc]].

\bibitem{Vainshtein:1972sx}
A.~I.~Vainshtein,
Phys. Lett. B \textbf{39} (1972), 393-394
doi:10.1016/0370-2693(72)90147-5

\bibitem{Nicolis:2008in}
A.~Nicolis, R.~Rattazzi and E.~Trincherini,
Phys. Rev. D \textbf{79} (2009), 064036
doi:10.1103/PhysRevD.79.064036
[arXiv:0811.2197 [hep-th]].


\bibitem{Afshordi:2006ad}
N.~Afshordi, D.~J.~H.~Chung and G.~Geshnizjani,
Phys. Rev. D \textbf{75} (2007), 083513
doi:10.1103/PhysRevD.75.083513
[arXiv:hep-th/0609150 [hep-th]].

\bibitem{Afshordi:2007yx}
N.~Afshordi, D.~J.~H.~Chung, M.~Doran and G.~Geshnizjani,
Phys. Rev. D \textbf{75} (2007), 123509
doi:10.1103/PhysRevD.75.123509
[arXiv:astro-ph/0702002 [astro-ph]].

\bibitem{Afshordi:2009tt}
N.~Afshordi,
Phys. Rev. D \textbf{80}, 081502 (2009)
doi:10.1103/PhysRevD.80.081502
[arXiv:0907.5201 [hep-th]].

\bibitem{Lin:2017oow}
C.~Lin and S.~Mukohyama,
JCAP \textbf{10} (2017), 033
doi:10.1088/1475-7516/2017/10/033
[arXiv:1708.03757 [gr-qc]].

\bibitem{Aoki:2018zcv}
K.~Aoki, C.~Lin and S.~Mukohyama,
Phys. Rev. D \textbf{98} (2018) no.4, 044022
doi:10.1103/PhysRevD.98.044022
[arXiv:1804.03902 [gr-qc]].

\bibitem{Aoki:2018brq}
K.~Aoki, A.~De Felice, C.~Lin, S.~Mukohyama and M.~Oliosi,
JCAP \textbf{01} (2019), 017
doi:10.1088/1475-7516/2019/01/017
[arXiv:1810.01047 [gr-qc]].

\bibitem{Mukohyama:2019unx}
S.~Mukohyama and K.~Noui,
JCAP \textbf{07} (2019), 049
doi:10.1088/1475-7516/2019/07/049
[arXiv:1905.02000 [gr-qc]].

\bibitem{DeFelice:2020eju}
A.~De Felice, A.~Doll and S.~Mukohyama,
JCAP \textbf{09} (2020), 034
doi:10.1088/1475-7516/2020/09/034
[arXiv:2004.12549 [gr-qc]].

\bibitem{Aoki:2020oqc}
K.~Aoki, A.~De Felice, S.~Mukohyama, K.~Noui, M.~Oliosi and M.~C.~Pookkillath,
Eur. Phys. J. C \textbf{80}, no.8, 708 (2020)
doi:10.1140/epjc/s10052-020-8291-1
[arXiv:2005.13972 [astro-ph.CO]].

\bibitem{Iyonaga:2018vnu}
A.~Iyonaga, K.~Takahashi and T.~Kobayashi,
JCAP \textbf{12} (2018), 002
doi:10.1088/1475-7516/2018/12/002
[arXiv:1809.10935 [gr-qc]].

\bibitem{Iyonaga:2020bmm}
A.~Iyonaga, K.~Takahashi and T.~Kobayashi,
JCAP \textbf{07} (2020), 004
doi:10.1088/1475-7516/2020/07/004
[arXiv:2003.01934 [gr-qc]].


\bibitem{Gleyzes:2014dya}
J.~Gleyzes, D.~Langlois, F.~Piazza and F.~Vernizzi,
Phys. Rev. Lett. \textbf{114} (2015) no.21, 211101
doi:10.1103/PhysRevLett.114.211101
[arXiv:1404.6495 [hep-th]].



\bibitem{deRham:2016ged}
C.~de Rham and H.~Motohashi,
Phys. Rev. D \textbf{95} (2017) no.6, 064008
doi:10.1103/PhysRevD.95.064008
[arXiv:1611.05038 [hep-th]].

\bibitem{Baker:2017hug}
T.~Baker, E.~Bellini, P.~G.~Ferreira, M.~Lagos, J.~Noller and I.~Sawicki,
Phys. Rev. Lett. \textbf{119} (2017) no.25, 251301
doi:10.1103/PhysRevLett.119.251301
[arXiv:1710.06394 [astro-ph.CO]].

\bibitem{Creminelli:2017sry}
P.~Creminelli and F.~Vernizzi,
Phys. Rev. Lett. \textbf{119} (2017) no.25, 251302
doi:10.1103/PhysRevLett.119.251302
[arXiv:1710.05877 [astro-ph.CO]].

\bibitem{Sakstein:2017xjx}
J.~Sakstein and B.~Jain,
Phys. Rev. Lett. \textbf{119} (2017) no.25, 251303
doi:10.1103/PhysRevLett.119.251303
[arXiv:1710.05893 [astro-ph.CO]].

\bibitem{Ezquiaga:2017ekz}
J.~M.~Ezquiaga and M.~Zumalac\'arregui,
Phys. Rev. Lett. \textbf{119} (2017) no.25, 251304
doi:10.1103/PhysRevLett.119.251304
[arXiv:1710.05901 [astro-ph.CO]].

\bibitem{Bean:2001wt}
R.~Bean, S.~H.~Hansen and A.~Melchiorri,
Phys. Rev. D \textbf{64} (2001), 103508
doi:10.1103/PhysRevD.64.103508
[arXiv:astro-ph/0104162 [astro-ph]].

\bibitem{Tsujikawa:2014mba}
S.~Tsujikawa,
Lect. Notes Phys. \textbf{892} (2015), 97-136
doi:10.1007/978-3-319-10070-8\_4
[arXiv:1404.2684 [gr-qc]].

\bibitem{Kase:2014yya}
R.~Kase and S.~Tsujikawa,
Phys. Rev. D \textbf{90} (2014), 044073
doi:10.1103/PhysRevD.90.044073
[arXiv:1407.0794 [hep-th]].

\bibitem{Kase:2014cwa}
R.~Kase and S.~Tsujikawa,
Int. J. Mod. Phys. D \textbf{23} (2015) no.13, 1443008
doi:10.1142/S0218271814430081
[arXiv:1409.1984 [hep-th]].

\bibitem{Lin:2014jga}
C.~Lin, S.~Mukohyama, R.~Namba and R.~Saitou,
JCAP \textbf{10} (2014), 071
doi:10.1088/1475-7516/2014/10/071
[arXiv:1408.0670 [hep-th]].

\bibitem{DeFelice:2011bh}
A.~De Felice and S.~Tsujikawa,
JCAP \textbf{02} (2012), 007
doi:10.1088/1475-7516/2012/02/007
[arXiv:1110.3878 [gr-qc]].

\bibitem{Gergely:2014rna}
L.~\'A.~Gergely and S.~Tsujikawa,
Phys. Rev. D \textbf{89} (2014) no.6, 064059
doi:10.1103/PhysRevD.89.064059
[arXiv:1402.0553 [hep-th]].

\bibitem{Gleyzes:2013ooa}
J.~Gleyzes, D.~Langlois, F.~Piazza and F.~Vernizzi,
JCAP \textbf{08} (2013), 025
doi:10.1088/1475-7516/2013/08/025
[arXiv:1304.4840 [hep-th]].

\bibitem{Aghanim:2018eyx}
N.~Aghanim \textit{et al.} [Planck],
Astron. Astrophys. \textbf{641} (2020), A6
doi:10.1051/0004-6361/201833910
[arXiv:1807.06209 [astro-ph.CO]].

\bibitem{Maeda:2022ozc}
K.~Maeda and S.~Panpanich,
Phys. Rev. D \textbf{105} (2022) no.10, 104022
doi:10.1103/PhysRevD.105.104022
[arXiv:2202.04908 [gr-qc]].
\end{thebibliography}
\end{document}